\documentclass{iaus}
\usepackage{graphicx}
\def\Msun{M_\odot}

\title[Planets around evolved stars] 
{A HET search for planets around evolved stars}

\author[Andrzej Niedzielski \& Alex Wolszczan]   
{Andrzej Niedzielski$^{1,2}$
 \and Alex  Wolszczan$^{2,1}$ }

\affiliation{$^1$Toru\'n Centre for Astronomy, Nicolaus Copernicus University \\ Gagarina 11,87-100 Toru\'n, Poland 
\\[\affilskip]
$^2$Dept. of Astronomy \& Astrophysics, The Pennsylvania State University, \\ 525 Davey Lab.oratory, University Park, 16802 PA, USA}

\pubyear{2008}
\volume{249}  
\pagerange{119--126}
\jname{Exoplanets: Detection, Formation \& Dynamics}
\editors{A.C. Editor, B.D. Editor \& C.E. Editor, eds.}
\begin{document}

\maketitle

\begin{abstract}
We present our ongoing survey of $\sim$1000 GK-giants with the 9.2-m Hobby-Eberly Telescope in search for planets around
evolved stars. The stars selected for this survey are brighter than 11 mag and are located in the section of the HR-diagram,
which is approximately delimited by the main sequence, the instability strip, and the coronal dividing line. We use the High
Resolution Spectrograph to obtain stellar spectra for radial velocity measurements with a 4-6 m s$^{-1}$ precision. So far,
the survey has discovered a planetary-mass companion to the K0-giant HD 17092, and it has produced a number of plausible
planet candidates around other stars. Together with other similar efforts, our program 
 provides information on planet formation around intermediate mass main sequence-progenitors and it will 
create the experimental basis with which to study dynamics of planetary systems around
evolving stars. 
\end{abstract}

\firstsection 
\section{Introduction}
Precision radial velocity (RV) studies have established more than a decade ago that
GK-giant stars exhibit RV variations ranging from days to many hundreds of days
(e.g. \cite[Walker et al. 1989]{walker89}, \cite[Hatzes \& Cochran 1993]{HC93}, \cite[Hatzes \& Cochran 1994]{1994ApJ...422..366H}). Enough observational
evidence has been accumulated to identify three distinct sources of this variability,
namely stellar pulsations, surface activity and a presence of substellar companions.
A possibility to discover planets around post-MS
giants, in numbers comparable to the current statistics of planets around MS-dwarfs
(e.g. \cite[Butler et al. 2006]{2006ApJ...646..505B}), offers a very attractive way to provide the much needed 
information on planet formation around intermediate mass MS-progenitors ($\geq 1.5\Msun$)
and to create a foundation for studies of the 
dynamics of planetary systems orbiting evolving stars (e.g. \cite[Duncan \& Lissauer 1998]{1998Icar..134..303D}).

Fourteen planet discoveries around GK-giants have been reported so far 
(\cite[Niedzielski et al. 2007]{2007ApJ...669.1354N}, and references
therein, \cite[Lovis \& Mayor 2007]{2007A&A...472..657L}, \cite[Johnson et al. 2007]{2007ApJ...665..785J}).
Locations of giants with planets in the HR diagram are shown in Fig. 1. All detections
have been made using the Doppler velocity technique with the RV precision ranging from
$\sim$5 to $\sim$25 m s$^{-1}$, exploiting the availability of many narrow absorption
features generated in the cool atmospheres of evolved stars. 
These developments demonstrate that sufficiently large surveys of post-MS giants should soon furnish enough
planet detections to meaningfully address the above problems. 

Initial analyses based on the currently available statistics
(\cite[Lovis \& Mayor 2007]{2007A&A...472..657L}, \cite[Johnson et al. 2007]{2007ApJ...665..785J}) suggest that the frequency
of massive planets is correlated with stellar mass. Because more massive
stars probably have more massive disks, these results appear to support
the core accretion scenarios of planet formation (\cite[Pollack et al. 1996]{1996Icar..124...62P}). Furthermore, \cite[Pasquini et al. 2007]{2007A&A...473..979P}  
have used the apparent lack of correlation between the frequency of planets
around giants and stellar metallicity to argue that this effect may imply
a pollution origin of the observed planet frequency - metallicity correlation
for main sequence stars (\cite[Fischer \& Valenti 2005]{2005ApJ...622.1102F}).
Finally, for planets around giants, the absence of planets on tight
orbits can be explained as the effect of post-MS evolution of their parent stars,
but, as discussed by \cite{2007ApJ...665..785J}, other scenarios must also be considered.
For example, the observed paucity of small orbital radii can be the result of faster
depletion of disks around more massive stars, as suggested by simulations
carried out by \cite{2007ApJ...660..845B} .

In this paper, we describe our contribution to searches for planets around post-MS stars with a
survey of $\sim$1000 GK-giants with the 9.2-m Hobby-Eberly Telescope. Our program has already discovered a number of
interesting planet candidates, first of which has been recently published by \cite{2007ApJ...669.1354N}.

\section{The survey}

\begin{figure}[b]
\begin{center}
\includegraphics[width=4.4in, angle=-90]{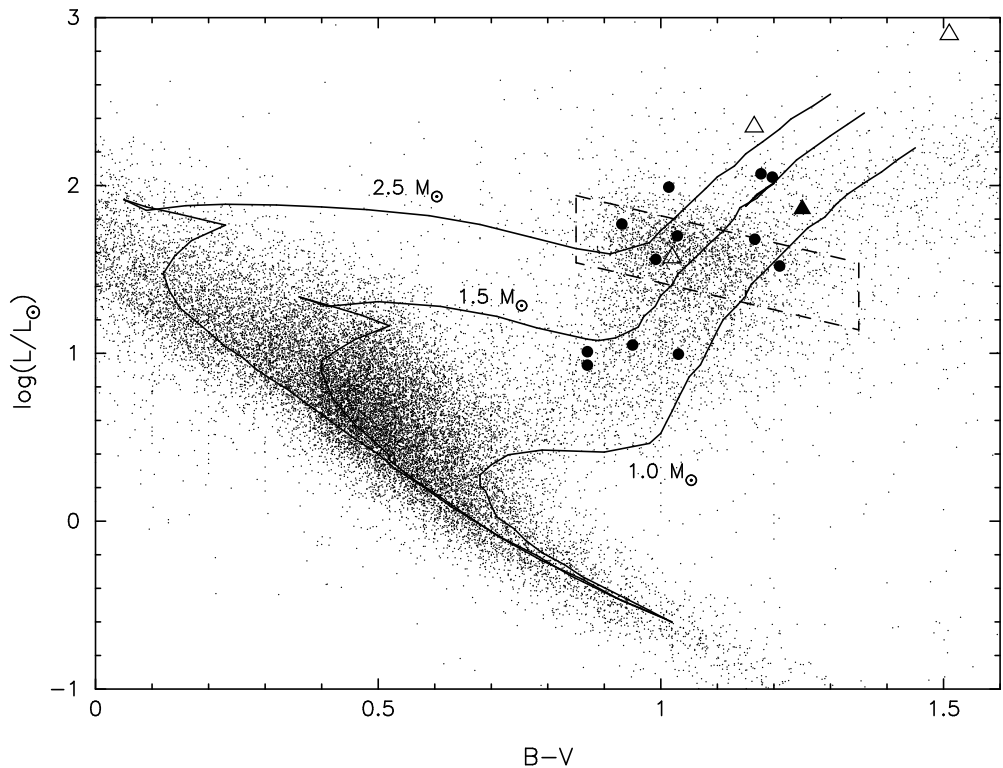} 
\caption{HR diagram for stars brighter than V=10 observable with the HET. The solid lines
are evolutionary tracks from \cite{1996A&AS..117..113G} for Z=0.02. The dashed box
delineates the red giant clump region (\cite[Jimenez et al. 1998]{jim98}). Symbols mark the stars with
published planet detections and planets from this survey discussed in the text.
 {\it (Filled triangle)} \cite{2007ApJ...669.1354N};
{\it (Open triangles)} unpublished detections from this survey; {\it (Filled circles)} other published
detections.
The diagram is based on data from the {\it Hipparcos} catalogue.}
\label{fig1}
\end{center}
\end{figure}

Our  long-term project to search for planets around evolved stars with the 9.2-m Hobby-Eberly Telescope (\cite[Ramsey et al. 1998]{lwr98}) and its High Resolution Spectrograph (\cite[Tull 1998]{tull98})
has been established in early 2004. The sample of stars we have been monitoring
is composed of two groups, approximately equal in numbers. The first one
falls in the ``clump giant'' region of the HR-diagram (\cite[Jimenez et al. 1998]{jim98}), which contains stars of various
masses over a range of evolutionary stages. The second group comprises stars, which have
recently left the MS and are located $\sim$1.5 mag above it. Generally, as shown in Fig. 1, all
our targets, a total of $\sim$1000 GK-giants brighter than $\sim$11 mag, occupy the area in
the HR-diagram, which is approximately defined by the MS, the instability strip, and the coronal
dividing line 
(a narrow strip in the HR-diagram marking the transition between stars with steady hot coronae and those with cool chromospheric winds \cite[Linsky \& Haisch 1979]{1979ApJ...229L..27L}).

The HET observations and data analysis for this survey have been described by \cite{2007ApJ...669.1354N}. Briefly, we observe with the HET
in its queue-scheduling mode and use the HRS at the R=60,000 resolution with the gas cell ($I_2$) inserted in the optical path.
In our target selection, we avoid bright objects, which are accessible to smaller telescopes. 
Consequently, more than 66\% of our target stars are fainter than V=8 mag.
The observing scheme follows the standard practices implemented in precision radial
velocity measurements with the iodine cell (\cite[Marcy \& Butler 1992] {iodine}). The spectral data 
used for RV measurements are extracted from the 17 echelle orders, which cover the 505 to 592 nm range of the $I_2$ cell spectrum.
The observing strategy consists of the initial set of  measurements of a
target star (2-3 exposures, typically 3-6 months apart), to check for any RV variability
exceeding a 30-50 m s$^{-1}$ threshold, followed by more frequent observations, if a significant variability is detected.
If the RV variability is confirmed, the star becomes part of the high priority list.

Radial velocities are measured by means of the commonly used $I_2$ cell calibration technique (\cite[Butler et al. 2006]{2006ApJ...646..505B}). 
A template spectrum is constructed from a high-resolution Fourier Transform Spectrometer (FTS) $I_2$ spectrum and a high signal-to-noise
stellar spectrum measured without the $I_2$ cell. 
Doppler shifts are derived from the least-squares fits of template spectra to
stellar spectra with the imprinted $I_2$ absorption lines.
The average radial velocity for each epoch is calculated as a mean value of
the independent determinations from the 17 usable echelle orders. The corresponding
uncertainties of these measurements are estimated assuming that errors obey
the Student's t-distribution. Typically, they fall in the 4-5 m s$^{-1}$ range at 1$\sigma$-level.  Radial velocities are referred to the Solar System barycenter using the \cite{1980A&AS...41....1S} algorithm, which is accurate enough given the RV precision limitations that are
intrinsic to the evolved stars.

As the intrinsic variability may contribute to the observed RV variations (e.g. \cite[Gray 2005]{2005PASP..117..711G}), 
stellar line profiles are studied in detail in search for any signatures of a rotation induced spot activity. Also the existing photometry databases like Hipparcos, Tycho or Northern Variability Sky Survey (\cite[Wo{\'z}niak et al. 2004]{2004AJ....127.2436W}) are used to study possible integrated light variations that might be interpreted as a result of pulsations. These analyses are reviewed  elsewhere (Niedzielski et al. this vol.).

\begin{figure}[b]
\begin{center}
\includegraphics[width=4.4in, angle=-90]{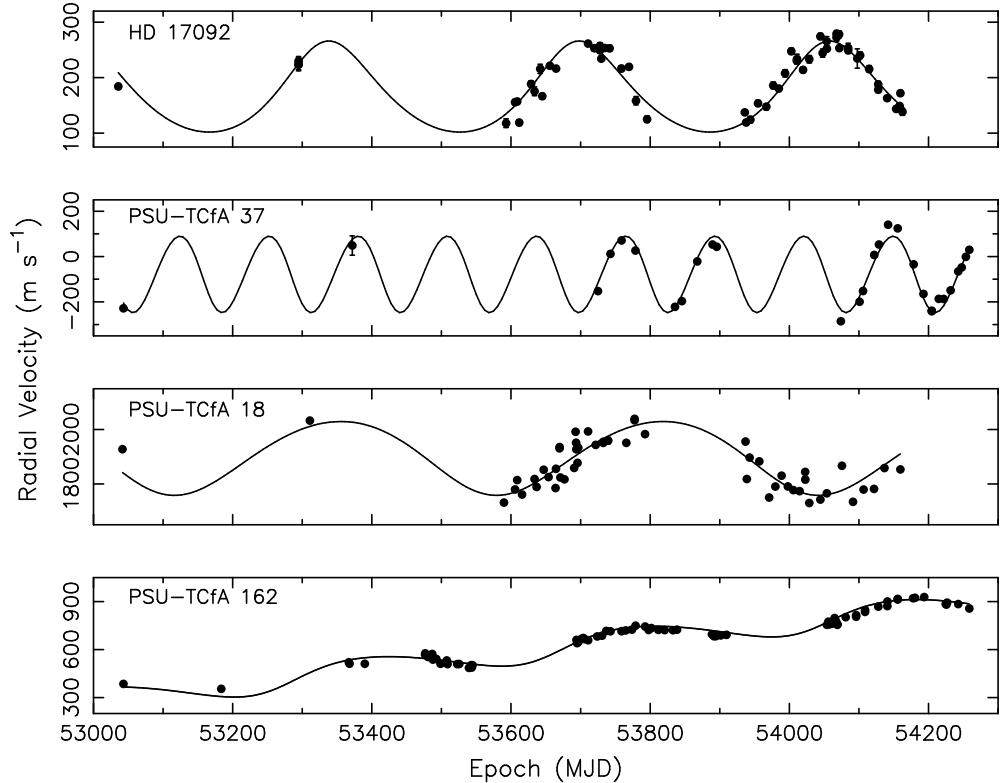} 
\caption{ Radial velocity measurements (filled circles) and the best-fit orbital
models (solid lines) for a sample of four K-giant stars monitored by the HET survey. For most data points, the circle size is larger than the sizes of
error bars.}
\label{fig2}
\end{center}
\end{figure}

\section{Results}
In almost four years of observations, we have obtained more than one RV measurement for
$>$600 GK-giant stars with a 4-6 m $s^{-1}$ precision.
Adopting a working definition of RV scatter $\le$40 m s$^{-1}$ for a stable (single) red giant, we find that
55 $\%$ of stars in that sample are single, 20 $\%$ are new binaries and 25$\%$ stars possibly have low-mass companions. 

We have been currently monitoring more than 30 planetary candidate companion stars 
and have obtained preliminary orbital solutions for most of them.
Each one obviously requires a thorough examination of stellar activity, which includes bisector analysis and a study of
H$_\alpha$ variations. The RV curves and the corresponding best-fit orbital models
for our first published planet around the K0 giant, HD 17092 (\cite[Niedzielski et al. 2007]{2007ApJ...669.1354N}),
and for three other examples of the detections that are being prepared for publication, are shown in Fig. 2.
The observed RV curves are highly repeatable and their
periods are not reproduced in the measured line bisector and photometric variations. Provisional
stellar mass estimates using \cite{2000A&AS..141..371G}  evolutionary tracks indicate a planetary
nature of the companions. It is quite clear that star 162 has a third, long-period  companion, whose nature
will be established in the course of further observations. The star 37 planet has
the most compact orbit among the existing detections (a = 0.6 AU), whereas the planet around
star 18 may be orbiting the most massive star in the existing sample (5.5 $\Msun$) and has an exceptionally high intrinsic RV noise.

A steadily increased number of stars observed in this survey makes it possible to carry out statistical studies of RV noise
properties of GK-giants. Our preliminary results  confirm the intrinsic RV jitter of red giants with the maximum of its
distribution at about 20 m s$^{-1}$. Furthermore, the RV scatter increases with B-V, easily reaching 100 m s$^{-1}$ for stars later than K5. Clearly, more observations are needed to understand the nature of the scatter, part of which may be contributed by short-period pulsations, which remain unresolved  by the sparse sampling of our survey.

Searches for planets around evolved stars are still in their infancy compared to
similar programs for solar-type stars, which have been steadily furnishing new
planet detections to bring the count up to over 250 at the time of this writing.
However, it is the former searches that are now needed to obtain new information
on stellar mass and time-dependent aspects of planet formation and evolution
that is not accessible through the latter ones. A continuation of the survey
described in this proposal, together with other similar programs, is already creating
a base of planet detections around GK-giants, which will soon become sufficient to
fully address the questions of stellar mass
and chemical composition dependence of planet formation
for masses $>1 \Msun$ and of the possible fates of planetary systems under
the influence of an evolving parent star. This knowledge will improve
our understanding of the astrophysics of planetary systems, it will provide
an experimental base for theories of the far future of the Solar System and it will
broaden our knowledge of
the astrophysical aspects of long-term survival of life on Earth and elsewhere,
including a possibility of the emergence of life on planets in the expanded
habitable zones of red giants (\cite[Lopez et al. 2005]{2005ApJ...627..974L}).


\acknowledgments
AN and AW were 
supported in part by the Polish Ministry of Science and Higher Education grant 1P03D 007 30.
AW also acknowledges a partial support from the NASA Astrobiology Program. 
The Hobby-Eberly Telescope (HET) is a joint project of the University of Texas at Austin, the Pennsylvania State University, Stanford University, Ludwig-Maximilians-Universit\"at M\"unchen, and Georg-August-Universit\"at G\"ottingen. The HET is named in honor of its principal benefactors, William P. Hobby and Robert E. Eberly.

\begin{discussion}

\discuss{Somebody}{Question.}{Answer. }

\end{discussion}


\begin{thebibliography}{}


\bibitem[Burkert \& Ida (2007)]{2007ApJ...660..845B} 
{Burkert, A., \& Ida, S.} 2007, \textit{Astrophysical Journal}, 660, 845 

\bibitem[Butler et al. (2006)]{2006ApJ...646..505B} 
{Butler, R.~P., et al.}  2006, \textit{Astrophysical Journal}, 646, 505 

\bibitem[Butler et al. (1996)]{Butler+1996} 
{Butler, R. P., Marcy, G. W., Williams, E., McCarthy, C., \& Dosanjh, P.} 1996, \textit{Publications of the Astronomical Society of the Pacific}, 	108, 500

\bibitem[Duncan \& Lissauer (1998)]{1998Icar..134..303D} 
{Duncan, M.~J., \& Lissauer, J.~J.} 1998, \textit{Icarus}, 134, 303 

\bibitem[Fischer \& Valenti (2005)]{2005ApJ...622.1102F} 
{Fischer, D.~A., \& Valenti, J.} 2005, \textit{Astrophysical Journal}, 622, 1102

\bibitem[Girardi et al. (1996)]{1996A&AS..117..113G} 
{Girardi, L., Bressan, A., Chiosi, C., Bertelli, G., \& Nasi, E.} 1996, \textit{Astronomy \& Astrphysics Supplement Series}, 117, 113 

\bibitem[Girardi et al.(2000)]{2000A&AS..141..371G} 
{Girardi, L., Bressan, A., Bertelli, G., \& Chiosi, C.}
 2000, \textit{Astronomy \& Astrphysics Supplement Series}, 141, 371 

\bibitem[Gray (2005)]{2005PASP..117..711G}
{Gray, D.~F.} 2005, \textit{Publications of the Astronomical Society of the Pacific}, 117,  711 

\bibitem[Hatzes \& Cochran (1993)]{HC93} 
{Hatzes, A.P., Cochran, W.D.} 1993, \textit{Astrophysical Journal}, 413, 339 

\bibitem[Hatzes \& Cochran (1994)]{1994ApJ...422..366H} 
{Hatzes, A.~P., \&  Cochran, W.~D.} 1994, \textit{Astrophysical Journal}, 422, 366 

\bibitem[Jimenez et al. (1998)]{jim98}
{Jimenez, R., Flynn, C. \& Kotoneva, E.} 1998, \textit{Monthly Notices of the Royal Astronomical Society}, 299, 515

\bibitem[Johnson et al. (2007)]{2007ApJ...665..785J} 
{Johnson, J.~A., et al.} 2007, \textit{Astrophysical Journal}, 665, 785 

\bibitem[Linsky \& Haisch (1979)]{1979ApJ...229L..27L} 
{Linsky, J.~L., \& Haisch, B.~M.} 1979, \textit{Astrophysical Journal Letters}, 229, L27 

\bibitem[Lopez et al. (2005)]{2005ApJ...627..974L} 
{Lopez, B., Schneider, J., \& Danchi, W.~C.} 2005, \textit{Astrophysical Journal}, 627, 974

\bibitem[Lovis \& Mayor (2007)]{2007A&A...472..657L} 
{Lovis, C., \& Mayor, M.} 2007, \textit{Astronomy \& Astrphysics}, 472, 657 

\bibitem [Marcy \& Butler (1992)] {iodine} 
{Marcy, G.W. \& Butler, R.~P.} 1992, \textit{Publications of the Astronomical Society of the Pacific}, 104, 270

\bibitem[Niedzielski et al. (2007)]{2007ApJ...669.1354N}
{Niedzielski, A., et al.} 2007, \textit{Astrophysical Journal}, 669, 1354 

\bibitem[Pasquini et al. (2007)]{2007A&A...473..979P} 
{Pasquini, L., D{\"o}llinger, M.~P., Weiss, A., Girardi, L., Chavero, C., Hatzes, A.~P., 
da Silva, L., \& Setiawan, J.} 2007, \textit{Astronomy \& Astrphysics}, 473, 979 

\bibitem[Pollack et al. (1996)]{1996Icar..124...62P} 
{Pollack, J.~B.,  Hubickyj, O., Bodenheimer, P., Lissauer, J.~J., Podolak, M., \& Greenzweig, Y.} 1996, \textit{Icarus}, 124, 62 

\bibitem[Ramsey et al. (1998)]{lwr98} 
{Ramsey, L.W., et al.} 1998, \textit{Proceedings of the SPIE}, 3352, 34

\bibitem[Stumpff (1980)]{1980A&AS...41....1S} 
{Stumpff, P.} 1980, \textit{Astronomy \& Astrphysics Supplement Series}, 41, 1 

\bibitem[Tull (1998)]{tull98} 
{Tull, R.G.} 1998, \textit{Proceedings of the SPIE}, 3355, 387

\bibitem[Walker et al. (1989)]{walker89} 
{Walker, G.A.H., Yang, S., Campbell, Bruce, I., Alan W.} 1989 \textit{Astrophysical Journal}, 343, 21

\bibitem[Wo{\'z}niak et al. (2004)]{2004AJ....127.2436W} 
{Wo{\'z}niak, P.~R., et al.} 2004, \textit{Astronomical Journal}, 127, 2436 


\end{thebibliography}
\end{document}